\documentclass[aps,prc,floatfix,nofootinbib,preprintnumbers,superscriptaddress,longbibliography,notitlepage,twocolumn]{revtex4-2}
\usepackage[dvipsnames,svgnames,x11names]{xcolor}
\usepackage{epsfig}
\usepackage{graphicx}
\usepackage{stmaryrd}
\usepackage{amssymb,tabularx,dcolumn}
\usepackage{supertabular,ltxtable}
\usepackage{longtable}
\usepackage{amsmath}
\usepackage{amstext}
\usepackage{float}
\usepackage{color}
\usepackage{bm}
\usepackage{hyperref}
\hypersetup{breaklinks=true,colorlinks=true,linkcolor=blue,citecolor=blue,filecolor=magenta,urlcolor=cyan}
\usepackage{mathtools}
\usepackage{multirow}
\usepackage{makecell}
\usepackage[utf8]{inputenc}
\usepackage{tabu}
\usepackage{braket}
\usepackage{subfigure} 
\usepackage{enumitem}
\usepackage[english]{babel}

\definecolor{pastelgray}{rgb}{0.81, 0.81, 0.77}
\definecolor{beaublue}{rgb}{0.9, 0.9, 0.93}

\newcommand{\midrule}{\hline}
\newcommand{\bottomrule}{\hline}

\begin{document}

\title{Predicting the neutrinoless double-beta decay matrix element of \texorpdfstring{$^{136}$}{}Xe using a statistical approach}

\author{M. Horoi}
\affiliation{Department of Physics, Central Michigan University, Mount Pleasant, MI 48859, USA}

\author{A. Neacsu}
\affiliation{International Center for Advanced Training and Research in Physics (CIFRA), Magurele, Romania}
\affiliation{Horia Hulubei National Institute of Physics and Nuclear Engineering (IFIN-HH), Magurele, Romania}

\author{S. Stoica}
\affiliation{International Center for Advanced Training and Research in Physics (CIFRA), Magurele, Romania}

\date{\today}

\begin{abstract}
 Calculation of the nuclear matrix elements (NMEs) for double-beta decay is of paramount importance for guiding experiments and for analyzing and interpreting the experimental data, especially for the search of the neutrinoless double beta decay mode ($0\nu\beta\beta$). However, there are currently still large differences between the NME values calculated by different methods, hence a quantification of their uncertainties is very much required.  In this paper we propose a statistical analysis of $0\nu\beta\beta$ NME for the $^{136}Xe$ isotope,  based on the interacting shell model, but using three independent effective Hamiltonians, emphasizing the range of the NMEs' most probable values and its correlations with observables that can be obtained from the existing nuclear data. Consequently, we propose a common probability distribution function for the $0\nu\beta\beta$ NME, which has a range of (1.55 - 2.65) at 90\% confidence level, with a mean value of 1.99 and a standard deviation of 0.37. 
\end{abstract}
\maketitle

\section{Introduction} \label{intro}

Double-beta decay (DBD) is an actively studied process due to its potential to provide insights into the nuclear structure of involved nuclei, the properties of neutrinos, and to test models beyond the Standard Model (SM)~\cite{Avignone2008,Vergados2012}.
Within the SM, this rare nuclear decay occurs with the emission of two electrons/positrons and two anti-neutrinos/neutrinos ($2\nu\beta\beta$), preserving the lepton number. However, the possibility of the decay occurring without the emission of neutrinos ($0\nu\beta\beta$), resulting in lepton number violation, is a highly intriguing theoretical possibility. In the case of neutrino exchange, this would imply that neutrinos are Majorana particles with non-zero mass, a feature beyond the original SM framework. While $2\nu\beta\beta$ transitions have been observed in eleven isotopes, no $0\nu\beta\beta$ transition has been detected yet. However, these transitions are actively sought in DBD experiments due to their potential to reveal phenomena beyond the SM.

The DBD half-life equations can be expressed, in a good approximation, as a product of some factors. Thus, the $2\nu\beta\beta$ half-life is a product of a phase space factor (PSF) describing the kinematics of the outgoing leptons~\cite{Doi1983,Doi1985,SuhonenCivitarese1998,Kotila2012, StoicaMirea2013, MireaPahomi2015}, and a nuclear matrix element (NME) describing the nuclear effects related to the nuclei involved in the decay. In the $0\nu\beta\beta$ half-life expression, besides the PSF and NME factors, an additional lepton number violation (LNV) factor appears as well, describing the particular BSM mechanism that may contribute to this decay mode. In principle, any LNV operator introduced in the Lagrangian can contribute, therefore the full half-life expression should be the sum of the individual contributions of all mechanisms and their interference terms~\cite{Doi1985,Vergados2012,Rodejohann2012,Deppisch2012,HoroiNeacsu2016prd,Neacsu2016ahep-dist,Ahmed2017}. 
In the absence of a signal indicating the $0\nu\beta\beta$ transition, the experimental lifetime limits and theoretical PSF and NME values are used to constrain the LNV parameters and associated BSM scenarios, typically under the assumption that only one mechanism is contributing at a time~\cite{18ho035502}.
Thus, progress in the DBD study needs the continuous improvement of the experimental set-ups and measurement techniques corroborated with accurate, reliable calculations of the theoretical quantities involved.
The current sensitivity of the DBD experiments reached limits of $10^{26}$ yr for the half-lives, and it is expected that the next  generation of experiments to push these limits to $10^{28}$ yr, thus covering the entire region of the neutrino inverted mass hierarchy~\cite{EngelMenendez2017,whiteppbb-2022}. The interpretation of these results in terms of values of the neutrino mass and constraints of the LNV parameters depend on reliable values of the calculated PSF and NME quantities.

\begin{center}
\begin{table*}[htb]
{\large
\begin{tabular}{lrrrrrrrrrrr}
\toprule
 Observable &  Data &  Error &  $svd_s$ &  $gcn_s$ &  $j5t_s$ &  $\mu_{svd}$ &  $\sigma_{svd}$ &  $\mu_{gcn}$ &  $\sigma_{gcn}$ &  $\mu_{j5t}$ &  $\sigma_{j5t}$ \\
\midrule
 $M_{0\nu}$ &   N/A &    N/A &    1.763 &    2.645 &    2.314 &        1.749 &           0.111 &        2.632 &           0.135 &        2.306 &           0.156 \\
 $M_{2\nu}$ & 0.018 &  0.001 &    0.025 &    0.069 &    0.060 &        0.022 &           0.003 &        0.061 &           0.007 &        0.052 &           0.007 \\
      P$GT$ & 0.150 &  0.021 &    0.163 &    0.545 &    0.512 &        0.141 &           0.059 &        0.457 &           0.105 &        0.333 &           0.220 \\
     P$BE2$ & 0.286 &  0.081 &    0.154 &    0.121 &    0.096 &        0.153 &           0.009 &        0.122 &           0.013 &        0.099 &           0.012 \\
 P$E_{2^+}$ & 1.313 &  0.150 &    1.498 &    1.363 &    1.513 &        1.494 &           0.089 &        1.352 &           0.089 &        1.507 &           0.098 \\
 P$E_{4^+}$ & 1.694 &  0.150 &    2.073 &    1.747 &    2.012 &        2.070 &           0.089 &        1.740 &           0.107 &        2.011 &           0.107 \\
 P$E_{6^+}$ & 1.892 &  0.150 &    2.178 &    1.892 &    2.254 &        2.192 &           0.088 &        1.884 &           0.125 &        2.212 &           0.096 \\
   POP$g_7$ & 2.930 &  0.100 &    2.705 &    2.716 &    3.143 &        2.702 &           0.187 &        2.705 &           0.209 &        3.082 &           0.267 \\
   POP$s_1$ & 0.057 &  0.006 &    0.089 &    0.025 &    0.020 &        0.090 &           0.018 &        0.025 &           0.006 &        0.021 &           0.006 \\
POP$h_{11}$ & 0.400 &  0.040 &    0.190 &    0.375 &    0.265 &        0.189 &           0.020 &        0.373 &           0.050 &        0.265 &           0.045 \\
     POP$d$ & 0.520 &  0.030 &    1.016 &    0.884 &    0.572 &        1.019 &           0.180 &        0.896 &           0.197 &        0.632 &           0.250 \\
      D$GT$ & 0.012 &  0.005 &    0.001 &    0.009 &    0.004 &        0.001 &           0.000 &        0.008 &           0.003 &        0.003 &           0.003 \\
     D$BE2$ & 0.413 &  0.011 &    0.342 &    0.194 &    0.158 &        0.337 &           0.023 &        0.195 &           0.026 &        0.163 &           0.028 \\
 D$E_{2^+}$ & 0.819 &  0.150 &    0.662 &    0.842 &    0.917 &        0.660 &           0.067 &        0.836 &           0.056 &        0.919 &           0.049 \\
 D$E_{4^+}$ & 1.867 &  0.150 &    1.389 &    1.873 &    2.113 &        1.403 &           0.131 &        1.861 &           0.116 &        2.087 &           0.082 \\
 D$E_{6^+}$ & 2.207 &  0.150 &    2.157 &    2.196 &    2.502 &        2.171 &           0.151 &        2.197 &           0.090 &        2.507 &           0.117 \\
   DVN$g_7$ & 0.000 &  0.150 &    0.102 &    0.174 &    0.130 &        0.100 &           0.010 &        0.172 &           0.014 &        0.132 &           0.023 \\
   DVN$s_1$ & 0.080 &  0.020 &    0.271 &    0.251 &    0.415 &        0.286 &           0.117 &        0.255 &           0.058 &        0.407 &           0.110 \\
DVN$h_{11}$ & 1.680 &  0.130 &    1.205 &    0.726 &    0.347 &        1.177 &           0.237 &        0.724 &           0.132 &        0.385 &           0.162 \\
     DVN$d$ & 0.240 &  0.050 &    0.423 &    0.850 &    1.108 &        0.437 &           0.132 &        0.850 &           0.118 &        1.076 &           0.158 \\
   DOP$g_7$ & 3.860 &  0.100 &    3.189 &    3.475 &    4.145 &        3.187 &           0.209 &        3.477 &           0.249 &        4.078 &           0.436 \\
   DOP$s_1$ & 0.200 &  0.020 &    0.263 &    0.083 &    0.049 &        0.264 &           0.047 &        0.084 &           0.020 &        0.052 &           0.017 \\
DOP$h_{11}$ & 0.620 &  0.060 &    0.264 &    0.658 &    0.625 &        0.269 &           0.049 &        0.658 &           0.093 &        0.613 &           0.121 \\
     DOP$d$ & 1.290 &  0.080 &    2.285 &    1.785 &    1.181 &        2.280 &           0.227 &        1.781 &           0.265 &        1.258 &           0.447 \\
\bottomrule
\end{tabular}

}
\label{allstats}
\caption{All relevant data and statistics for all selected observables. See section \ref{results} for notations and details.}
\end{table*}
\end{center}

The progress of the theoretical methods for relativistic wave function computations, now provides PSF calculations with a high degree of confidence for all the double-beta decay modes and transitions~\cite{Kotila2012, StoicaMirea2013, MireaPahomi2015}. However, the same level of confidence is not yet valid for the NME calculation, which still remain the main source of uncertainty for the DBD lifetime.
There are several nuclear structure methods for the NME calculation, the most used being: interacting shell model methods~\cite{Caurier1990,Caurier1996, Caurier2005, HoroiStoicaBrown2007, HoroiStoica2010, Horoi2013, HoroiBrown2013, SenkovHoroi2014, NeacsuHoroi2015, NeacsuHoroi2016,18ho035502}, pn-QRPA methods~\cite{SuhonenCivitarese1998, Simkovic1999, Stoica2001, Rodin2006, KortelainenSuhonnen2007, Faessler2012, SimkovicRodin2013}, IBA methods~\cite{Barea2009, Barea2013}, Energy Density Functional method~\cite{Rodriguez2010}, PHFB~\cite{Rath2013}, Coupled-Cluster method (CC)~~\cite{Novario2021}, in-medium generator coordinate method (IM-GCM)~\cite{Yao2020}, and valence-space in- medium similarity renormalization group method (VS-IMSRG)~\cite{Belley2021}. Each of these methods have their strengths and weakness, largely discussed over time in the literature, and the current situation is that there are still significant differences between NME values calculated with different methods, and sometimes, even between NME values calculated with the same methods (see for example the review~\cite{EngelMenendez2017}). For the $2\nu\beta\beta$ decay NMEs are products of two Gamow-Teller (GT) transition amplitudes, and most of the nuclear methods overestimate them, in comparison with experiment. This drawback is often treated by introducing a quenching factor that multiplies the GT matrix element and reduces its strength. This procedure is viewed as equivalent to using a quenched axial vector constant, instead of its bare value $g_A$ = 1.27. The NME calculation for $0\nu\beta\beta$ decay is more complicated, since besides the GT transitions, other transitions may contribute as well. Also, the NME values calculated by different methods may differ by factors of 3-4 for most relevant isotopes, including $^{136}Xe$ (see e.g. Fig. 5 of Ref.~\cite{EngelMenendez2017}, and Refs.~\cite{Yao2020,Novario2021}). The uncertainties in the NME values are further amplified when predicting half-lives, since they enter at the power of two in the inverse lifetime formula. In addition, given that there is no measured lifetime for this decay mode to compared with, these uncertainties in the calculated NME affect the prediction and interpretation of the existing $0\nu\beta\beta$ half-life limits and the planning of performances for the future DBD experiments.

Among the nuclear methods for calculating NMEs, the shell model based methods have some advantages, such as the inclusion of all correlations between nucleons around the Fermi surface, preserving all symmetries of the nuclear many-body problem, and the use of nucleon-nucleon (NN) interactions tested for other observables and for different mass regions of nuclei. The construction and use of effective NN Hamiltonians in accordance with the model spaces is a key ingredient in calculations. Therefore, one question that arises is the stability of the calculated NME values to small changes in the parameters of effective Hamiltonians.
In a previous recent paper~\cite{Horoi-prc22}, we presented a statistical analysis of the NME distribution for $^{48}Ca$  to random changes of two body matrix elements (TBME) calculated with shell model methods in a fp model space with three different effective Hamiltonians, namely FPD6, GXPF1A and KB3B. Besides the stability of NME to these changes, we also investigated the correlation between the changes in the $0\nu\beta\beta$ NME and the changes in other observables, such as  $2\nu\beta\beta$ NME, GT strengths, B(E2) transition probabilities, excited states energies, occupation probabilities, etc. Based on this statistical analysis with the three Hamiltonians, we proposed  a common probability distribution function for $0\nu\beta\beta$ NME which has a range of (0.45 – 0.95) at $90\%$ confidence level with a mean value of 0.68~\cite{Horoi-prc22}.  A similar analysis for $^{76}Ge$ using ab-initio nuclear methods, although with a smaller number of observables and a much smaller statistics, was recently presented in Ref.~\cite{Belley-2022}. Indeed, it is important to provide uncertainty quantification for observables of physical processes like $0\nu\beta\beta$ NME where experimental data for verification is limited.

\begin{figure*}[htb]
\includegraphics[clip, trim={2.5cm 3.5cm 1.5cm 4.0cm},width=20cm]{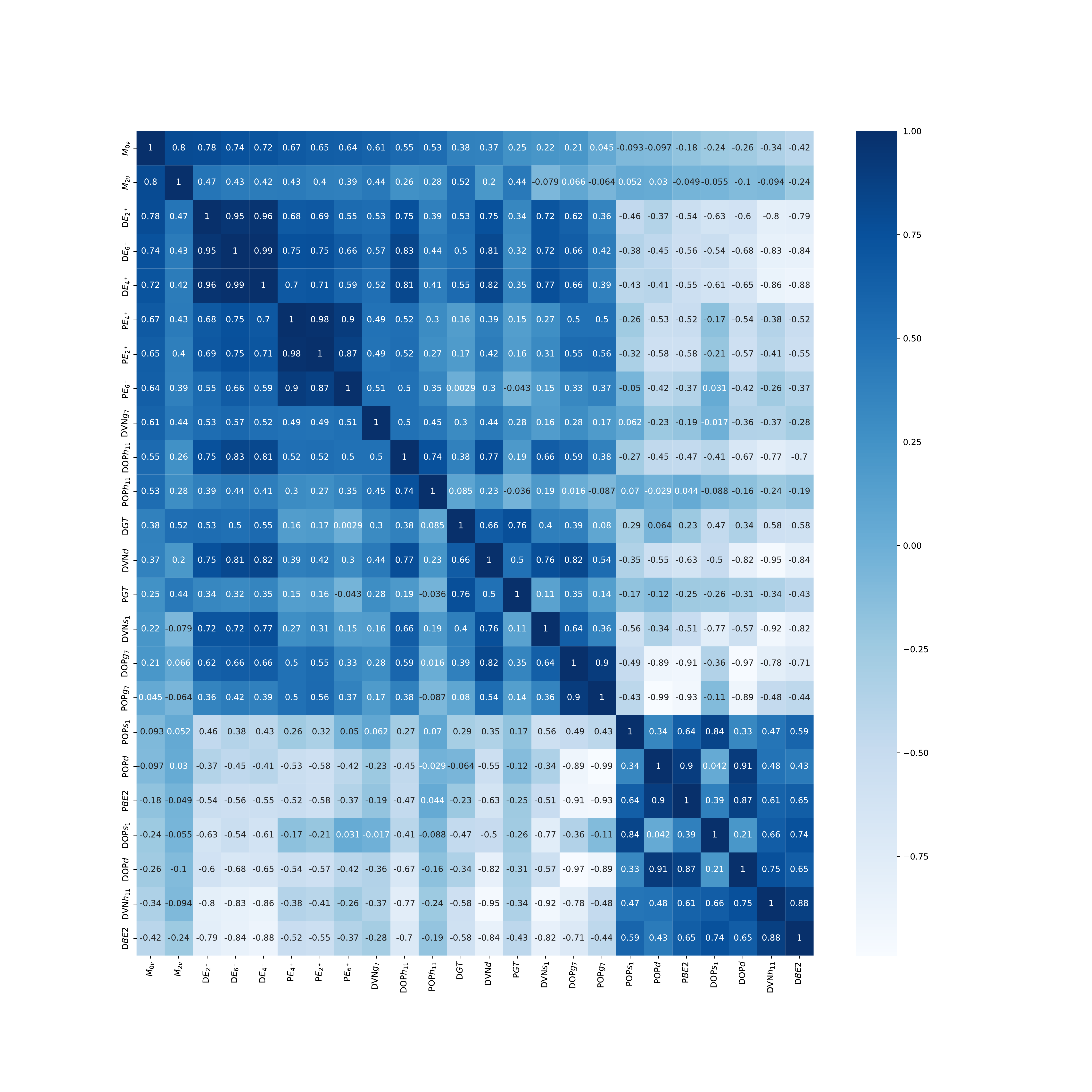}
\caption{The heat-map for all 24 observables when using the SVD effective Hamiltonian. See section \ref{results} for notations and analysis.}
\label{svd-hm}
\end{figure*}

In this paper we propose a similar statistical analysis of $0\nu\beta\beta$ NME for $^{136}Xe$, which is theoretically among the most suitable for NME calculation using a shell model approach, and it is also among the most promising isotope for experimental search of $0\nu\beta\beta$ transitions. We only consider in this work the standard light left-handed (LH) neutrino exchange mass mechanism, which is presently viewed as the most likely to contribute to the $0\nu\beta\beta$ decay process. The calculations are performed using three independent effective Hamiltonians SVD~\cite{Chong2012}, jj55t~ \cite{HoroiBrown2013} and GCN5082~\cite{GCN-int}, for the jj55 model space that is appropriate for $^{136}Xe$. These effective Hamiltonians are obtained starting with a theoretical Bruekner G-Matrix effective Hamiltonians that are further fine-tuned to describe the experimental energy levels for a reasonably large number of nuclei that can be investigated in the corresponding model spaces. These effective Hamiltonians are described by a small number of single particle energies and a finite number of two-body matrix elements. As a by-product, the wave functions produced by these Hamiltonians can be used to describe and predict observables, such as the electromagnetic transition probabilities, Gamow-Teller transitions probabilities, nucleon occupation probabilities, spectroscopic factors, etc, using relatively simple changes of the transition operators in terms of effective charges and quenching factors. These quantities are calibrated to the existing data. For $0\nu\beta\beta$ NMEs such calibrations are not yet possible due to the lack of experimental data confirming the transition. However, different existing effective Hamiltonians for nuclei involved in a given $0\nu\beta\beta$  decay produce smaller ranges of the NME. In addition, some recent ab-initio methods, such as IM-SRG~\cite{Yao2020,Belley2021}, built on the modern advances in the shell model by providing ab-initio derived effective Hamiltonians and effective transition operators can provide some guidance for calibrating the shell model calculations for $0\nu\beta\beta$ NMEs.

\begin{figure*}[t]
\includegraphics[clip, trim={0.2cm 0.2cm 0.0cm 0.0cm},width=16.0cm]{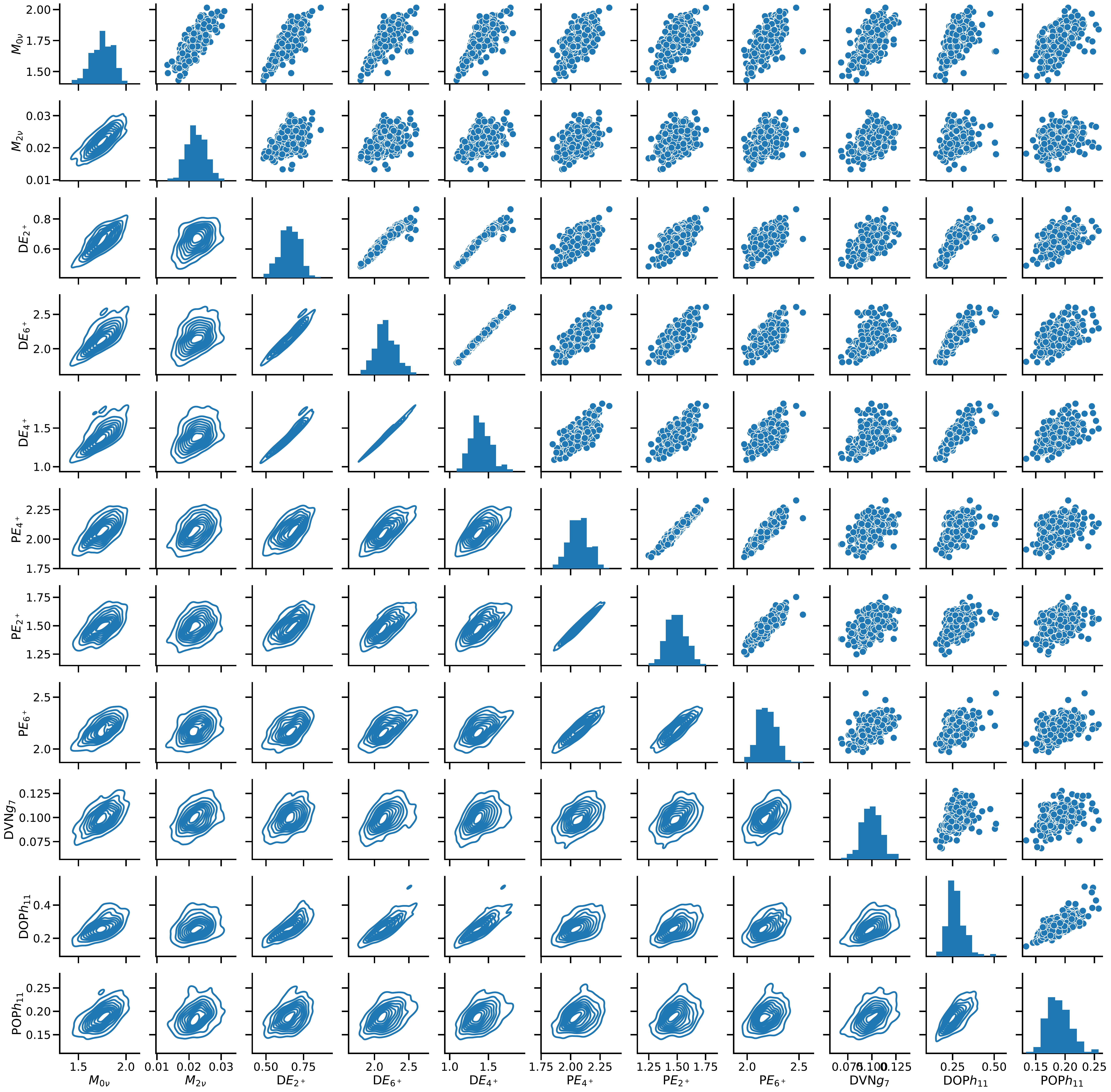}
\caption{Correlation matrix for observables that have correlation factor greater than 0.5, when using the SVD Hamiltonian. See section \ref{results} for notations and analysis.}
\label{svd-corr}
\end{figure*}

Following the analysis line from~\cite{Horoi-prc22}, we study the robustness of the $^{136}$Xe $0\nu\beta\beta$ NME values to small changes of the parameters of the above mentioned Hamiltonians, and also examine the correlation between the changes in $0\nu\beta\beta$ NMEs and other observables, for which the experimental data exists. Furthermore,  we investigate the range of possible $0\nu\beta\beta$ NME values and their correlations with several observables that can be extracted from the existing nuclear data. Finally, using a statistical analysis based on the Bayesian Averaging Model~\cite{Kejzlar_2020,Cirigliano_2022}, we propose a common probability function for the $0\nu\beta\beta$ NME, a plausible range, its expectation value and its uncertainty. The application of the Bayesian Averaging Model here is novel; it was not possible in the $^{48}$Ca case due to lack of relevant experimental data that was highly correlated with $0\nu\beta\beta$ NME.

The paper is organized as follows. In section II the calculation methods of the observables and the statistical model are presented.
Then, in section III we present the results and discussions on their relevance, followed by an statistical analysis based on the Bayesian Model Averaging method in section IV, and in section V we end with conclusions and outlook.

\section{The Model} \label{model}

Following the analysis available in Ref.~\cite{Horoi-prc22}, we extend our study to the $^{136}$Xe isotope that actively investigated or proposed in several leading current and future DBD experiments \cite{nEXO_2022, Kamland-Zen-2022, PandaX-47-2022, XENONnT-2022, Darwin-2020}. In this work we are also concerned calculating the $0\nu\beta\beta$ NME starting from three popular shell model effective Hamiltonians alongside several observables that can be compared to their experimental values.

The calculations reported here are done within the interacting shell model in the $jj55$ model space consisting of the $0g_{7/2}, 1d_{5/2}, 1d_{3/2}, 2s_{1/2}$ and $0h_{11/2}$ orbitals that assumes $^{100}$Sn as a core, covering the sector of the nuclear chart between N,Z=50 and N,Z=82. One concern regarding the $jj55$ model space is the missing Gamow-Teller strengths when compared to the calculated Ikeda sum-rule. This is attributed to missing spin-orbit partner orbitals of $0g_{7/2}$ and $0h_{11/2}$. Calculations in $jj77$ model space are too complex~\cite{HoroiBrown2013} and not presently suitable for a statistical analysis.

As starting effective Hamiltonians we use SVD~\cite{Chong2012}, jj55t \cite{HoroiBrown2013}, and GCN5082~\cite{GCN-int}. The jj55t effective Hamiltonian (also known as sn100t~\cite{REBEIRO2020135702}) is very similar to sn100pn~\cite{PhysRevC.71.044317}, except with minor modifications and was used in Refs.~\cite{HoroiBrown2013,REBEIRO2020135702} to calculate NMEs for $^{136}$Xe $0\nu\beta\beta$ decay. The GCN5082 effective Hamiltonian is based on a renormalized G-matrix~\cite{Jensen1995} obtained from the Bonn-C  nucleon-nucleon potential ~\cite{Machleidt2001}. The final effective Hamiltonian was obtained through a mostly monopole fit to about 300 energy levels from approximately 90 nuclei in the region with a root mean-square (RMS) deviation of about 150 keV~\cite{GCN-int}. 
 Similarly to GCN5082, SVD~\cite{Chong2012} also starts with a Bonn-C potential and renormalized via a G-matrix method for the core polarization effects~\cite{Hjorth-Jensoen-RevModPhys-2003}. The $T = 1$ monopoles and the single-particle energies where obtained by fitting to the binding energies of 157 experimentally measured~\cite{AME2003%,NNDC2} 
} low-lying yrast states in $^{102-132}$Sn nuclei.
These three Hamiltonians are further modified by introducing random perturbations within the range of $\pm 10\% $ to their two-body matrix elements (TBME) with the aim of getting a range to the shell model $0\nu\beta\beta$ NME values and the correlations between the $0\nu\beta\beta$ NME and the other observables. 
For the purpose of this study, we generate 1000 effective Hamiltonians via random perturbations from each starting Hamiltonian. Just like in the case of $^{48}$Ca~\cite{Horoi-prc22}, the single-particle energies were kept unmodified, as not to interfere with the magicity of the $^{100}$Sn core.

The aim of this study is to explore the relationship between the $0\nu\beta\beta$ NME and other measurable observables for each starting effective Hamiltonian. The research also aims to establish theoretical limits for each observable, examine the shape of different distributions for each observable and starting Hamiltonian, use this data to determine the impact of different starting Hamiltonians on the most favorable distribution of the $0\nu\beta\beta$ NME, and ultimately identify the most favorable value of the $0\nu\beta\beta$ NME and its estimated theoretical uncertainty.

The observables that we calculate and compare to their experimental values are: $2\nu\beta\beta$ NME, the energies of the first $2^+$, $4^+$, and $6^+$ states in the parent ($^{136}$Xe) and daughter ($^{136}$Ba) nuclei, B(E2)$\uparrow$ transition probabilities for $^{136}$Xe and $^{136}$Ba to the first $2^+$ states, the Gamow-Teller transition probability for the transition from $^{136}$Xe and from $^{136}$Ba to the $1^+$ excited state in $^{136}$Cs, and the neutron and proton occupancies for $^{136}$Xe and $^{136}$Ba above the $^{100}$Sn core in the jj55 model space shells. The number of observables that we calculate for each sample is 24, including the $0\nu\beta\beta$ NME.

Other observables related to double-beta decay, such as one-muon capture (OMC) rates, have also been studied in the literature~\cite{Zinlatulina2019}. Some recent references for OMC analyses can be found in Refs.~\cite{Siiskonen1998} and~\cite{Fox-2020}. However, the calculation of the OMC rates is quite complex, as it depends on multiple factors that contribute to the decay amplitude, which can lead to uncontrollable interference effects~\cite{Zinlatulina2019}. Additionally, it is highly sensitive to the effective Hamiltonian used~\cite{Cirigliano-PhysRevLett-2018}. Due to this complexity, we have decided not to include the OMC rates to our list of observables in this study.

The $0\nu\beta\beta$ NME is related to the half-life of the respective process \cite{HoroiStoica2010}, where we only consider the contribution from the light left-handed neutrino exchange mechanism, which is likely to contribute to the $0\nu\beta\beta$ decay. The methodology of calculating the $0\nu\beta\beta$ NME, $M_{0\nu}$,  within the shell model was extensively described elsewhere \cite{HoroiStoica2010,Horoi2013,18ho035502} and it will not be repeated here (see also Eq. (1) of Ref. \cite{Horoi-prc22}). It includes a short range correlation function that can be viewed as an effective modification of the bare operator. In Ref. \cite{Horoi-prc22} we were able to select a short-range correlation function based on comparisons with similar calculations with ab-initio effective operators. Unfortunately, such a comparison is not possible for $^{136}$Xe, while no such ab-initio calculations are available. Therefore, we choose a short-range correlation function based on the widely utilized CD-Bonn parametrization (see e.g. \cite{NeacsuHoroi2015,NeacsuHoroi2016,18ho035502}). One should also add that as in Ref. \cite{Horoi-prc22}, here we also use the closure approximation. It is well known that the dependence of the closure energy is very mild, and although one can find optimal closure energies for each Hamiltonian~\cite{SenkovHoroi2013,SenkovHoroi2014,SenkovHoroiBrown2014,Senkov2016}, here we use the same closure energy of 3.5 MeV~\cite{NeacsuHoroi2015} in all cases.

\begin{figure*} 
\includegraphics[clip, trim={2.5cm 5.0cm 2.5cm 6.2cm}, width=13cm]{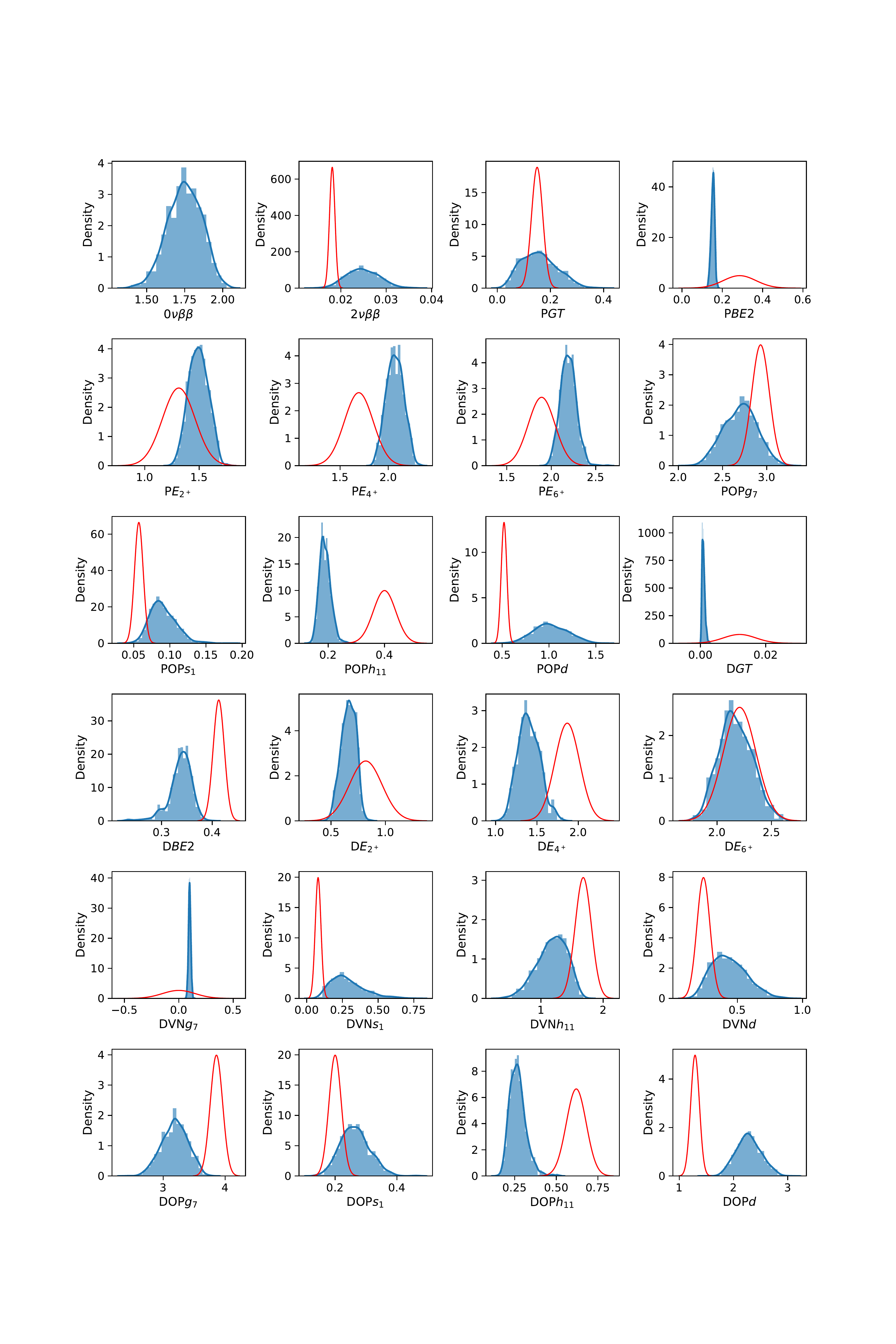}
\caption{Distributions based on experimental data in red compared with the those obtained from the SVD starting Hamiltonian.}
\label{svd-diag-dist}
\end{figure*}

The $2\nu\beta\beta$ NME squared is proportional to the inverse half-life of the respective process \cite{HoroiStoicaBrown2007} (see also Eq. (2) of Ref. \cite{Horoi-prc22}). The $2\nu\beta\beta$ NME, $M_{2\nu}$, can be calculated with
\begin{equation}
  M_{2\nu} = \sum_k \frac{q\left<0^+_f \mid \sigma \tau^- \mid 1^+_k\right>\left<1^+_k \mid \sigma \tau^- \mid 0^+_i\right>q}{E_k-E_0} \ ,
\label{m2n}
\end{equation}
where the summation is on the $1^+_k$ states in $^{136}$Cs,  $E_0=Q_{\beta\beta}/2+\Delta M(^{136}$Sc$-^{136}$Xe), and $q$ is the quenching factor of the GT matrix element.

In our analysis of the shell model for $2\nu\beta\beta$ NME values, we maintain consistency by utilizing the same quenching factor of $q=0.70$ for both the NME and GT strengths. Additionally, we maintain the standard canonical values for effective charges in our calculations of B(E2)$\uparrow$. Other observables, including excited state energies, GT strengths to the first $1^+$ state of $^{136}$Cs, B(E2)$\uparrow$ to the first $2^+$ state in the parent and daughter nuclei, and s.p. occupation probabilities, are calculated using the established shell model methodology. Here we use in all cases the same effective charges ($e_p=1.5$ and $e_{n}=0.5)$) for the B(E2)$\uparrow$, and the same quenching factor ($q=0.70$) for the the GT strengths and $M_{2\nu}$.

\section{Results of the Statistical Analysis} \label{results}

The experimental data used in this study is listed in Table~\ref{allstats}. The excitation energies of the $2^+$, $4^+$ and $6^+$ states of $^{136}$Xe and $^{136}$Ba in MeV are taken from Ref.~\cite{NNDC-MCCUTCHAN2018}. The $2\nu\beta\beta$ NME (in MeV$^{-1}$) is from Ref.~\cite{Barabash2020}.  B(E2)$\uparrow$ electric quadrupole transition probabilities (in $e^2b^2$) come from Ref.~\cite{Pritychenko2016}. The Gamow-Teller transition probabilities to the first excited $1^+$ state in $^{136}$Cs are from Ref.~\cite{Frekers2013}. $^{136}$Ba neutron vacancies are taken from  Ref.~\cite{136Xe-Ba-neutron-occupations-2016}, while $^{136}$Xe and $^{136}$Ba proton occupancies are from Ref.~\cite{136Xe-Ba-proton-occupations-20216} . The experimental errors for the excitation energies are very small, and for the calculations we use the typical theoretical RMS value of 150 keV~\cite{Honma2004}. All observables have experimental data available, except for the $0\nu\beta\beta$ NME. Therefore, a statistical analysis of the $0\nu\beta\beta$ NME is performed using the Bayesian Averaging Model (see Section~\ref{statistical-inference}).

The primary outcomes of this study are presented in Table~\ref{allstats} and Figures~\ref{svd-hm}-\ref{svd-diag-dist}. Here, the "parent nucleus" stands for $^{136}$Xe, "daughter nucleus" stands for $^{136}$Ba, and "intermediary nucleus" stands for $^{136}$Cs from the perspective of $\beta\beta$ transitions. When presenting the results referring to one single nucleus, we use the the letter "P" at the beginning of a label for an observable indicates that it is related to the parent nucleus, while the letter "D" denotes observables for the daughter nucleus. In the table and figures $M_{0\nu}$ are the $0\nu\beta\beta$ NMEs, and $M_{2\nu}$ denote the $2\nu\beta\beta$ NMEs. With P$GT$ and D$GT$ we present the Gamow-Teller strengths to the first excited $1^+$ state in the $^{136}$Cs intermediate nucleus from $^{136}$Xe and from $^{136}$Ba, respectively. P$B(E2)\uparrow$ and D$B(E2)\uparrow$ are the electric quadrupole transition probabilities ($0^+ \rightarrow 2^+$) for $^{136}$Xe and $^{136}$Ba, respectively. P$E_{2^+}$, P$E_{4^+}$, P$E_{6^+}$ and D$E_{2^+}$, D$E_{4^+}$, and D$E_{6^+}$ denote the energy of the first $2^+$, $4^+$, and $6^+$ excited states, for $^{136}$Xe and $^{136}$Ba and respectively. POP${g_7}$, POP${s_1}$, POP${h_1}$, and POP$_d$ stand for the proton occupation probabilities of the $0g_{7/2}, 2s_{1/2}$, $0h_{11/2}$, and $d$ orbitals in $^{136}$Xe, while DOP${g_7}$, DOP${s_1}$, DOP${h_1}$, and DOP$_d$ are the proton occupation probabilities of the $0g_{7/2}, 2s_{1/2}$, $0h_{11/2}$, and $d$ orbitals in $^{136}$Ba. DVN${g_7}$, DVN${s_1}$, DVN${h_1}$, and DVN$_d$ represent the neutron vacancy probabilities in $^{136}$Ba. The experimental proton occupancies do not distinguish between the $1d_{5/2}$ and $1d_{3/2}$ orbitals, thus we add our results for both orbitals into a single proton occupation probability.

In the columns of Table~\ref{allstats}  we show from left to right the experimental values (Data), the adopted experimental errors (Error), the calculated values of the observables using the starting Hamiltonians SVD (labeled "$svs_s$"), GCN5082 (labeled "$gcn_s$") and jj55t (labeled "$j5t_s$"), the mean value obtained after 1000 samples for each starting Hamiltonian, denoted with $\mu$, followed by the standard deviation $\sigma$.
Overall, one can see from Table~\ref{allstats}  that the SVD starting Hamiltonian produces $M_{2\nu}$ NMEs that are closest to the experimental value, thus needing the least amount of quenching when compared to those of GCN5082 or jj55t. 
For the P$GT$ and D$GT$ one observes that the SVD results are closest to the experimental data for the parent nucleus, overestimating the result by much less than GCN5082 and jj55t. However, for the daughter's GT, GCN5082 was best, with SVD underestimating the result the most.
For P$B(E2)\uparrow$ and D$B(E2)\uparrow$ SVD shows values closest to the experiment. Carefully adjusting the values of the effective charges would improve the results for all three Hamiltonians, but that is beyond the scope of our analysis. 
The excitation energies are better described by GCN5082, in large part because the  GCN5082 starting Hamiltonian was fine-tuned with data for more nuclei and energy levels than SVD and jj55t. 
Overall, GCN5082 appears to describe the occupations and vacancies best.

Figures~\ref{svd-hm}-\ref{svd-diag-dist} present more detailed statistical results obtained with the SVD starting Hamiltonian. The corresponding figures for GCN5082 and jj55t starting effective Hamiltonians look similar and are not included here.
Figure~\ref{svd-hm} presents the complete correlation matrix for all 24 observables that we calculate, with the number values denoting the Pearson coefficient R.
The color intensity highlights the value of Pearson coefficient R between -1 as white and 1 as dark blue. For ease of use, a color scale is also shown on the right side. The lines are listed in descending order for the value of the Pearson coefficient R of an observable and $M_{0\nu}$. This figure is particularly interesting because it reveals the correlations between all of the observables, not just related to $M_{0\nu}$. 

Figure~\ref{svd-corr} illustrates the more interesting cases for correlations between the observables, where the value of the Pearson coefficient R of an observable and $M_{0\nu}$ is higher than 0.5. Since on the diagonal every observable would correlate perfectly with itself, we utilize that space to plot the histograms for the probability distributions. On top of the diagonal we present scatter plots for pairs of observables forming the coordinates with  a reduced set of data points for ease of viewing, while below the diagonal we show contour plots emphasising the density of points considering all the available data. Visually, higher values of the Pearson correlation coefficient R result in scatter plots and contour plots clustering closer to a diagonal line in each graph. Most noticeable examples include the energy levels that correlate with each other and, significantly for this study, the 
$0\nu\beta\beta$ NME and the $2\nu\beta\beta$ NME with R=0.8 in the case of the SVD Hamiltonian.

 Figure~\ref{svd-diag-dist} details the histograms of the 24 observables with increased detail of the data bins and adding the experimental data in the form of a Gaussian distribution displayed with a red curve. This Gaussian was obtained with the experimental values providing the mean and its width constrained by the experimental error for the standard deviation.
 Encasing the probability distributions with a blue line is the kernel-density estimate usedfor the analysis detailed in Section~\ref{statistical-inference}.

Interestingly, the correlations between the $0\nu\beta\beta$ NME and the strengths of the parent and daughter Gamow-Teller transitions to the first $1^+$ state in $^{136}$Cs are significantly reduced, while the correlation with the $2\nu\beta\beta$ NME is very strong. One explanation for this phenomenon could be related to the fact that the product of the GT matrix elements describing transitions to the first $1^+$ state in $^{136}$Cs in Eq. (\ref{m2n}) does not significantly contribute to the total sum of all excited $1^+$ states in the intermediate nucleus.

Other observables that have relatively high correlations with the $0\nu\beta\beta$ NME (detailed in Fig.~\ref{svd-corr}) are the energies of the $2^+$, $4^+$ and $6^+$ states in both $^{136}$Xe and $^{136}$Ba with the correlators $R$ between 0.64 and 0.78.
The ${g_{7/2}}$ neutron vacancies in $^{136}$Ba correlate with the $0\nu\beta\beta$ NME at $R=0.61$. The proton occupancies of the ${h_{11/2}}$ orbital in $^{136}$Ba correlate with the $0\nu\beta\beta$ NME at $R=0.55$, while the proton occupancies of the ${h_{11/2}}$ orbital in $^{136}$Xe correlate with the $0\nu\beta\beta$ NME at $R=0.53$.

From the full correlation matrix in Figure~\ref{svd-hm}, we notice the B(E2)$\uparrow$ of the $^{136}$Xe and $^{136}$Ba cases. P$BE2$ correlates very strongly with POP$d$ ($R=0.9$) and DOP$d$ ($R=0.91$), while it anti-correlates significantly with DOP$g_7$ ($R=-0.91$) and POP$g_7$ ($R=-0.93$). P$BE2$ also shows reasonable anti-correlations with D$E_{2^+}$ ($R=-0.54$), D$E_{4^+}$ ($R=-0.55$), D$E_{6^+}$ ($R=-0.56$), P$E_{2^+}$ ($R=-0.58$) and D$E_{4^+}$ ($R=-0.52$). For 
P$BE2$ (right-most column of Fig.~\ref{svd-hm}) we highlight the correlations with DOP$s_1$ ($R=0.74$) and DVN$h_{11}$ ($R=0.88$), anti-correlating with DVN$d$ ($R=-0.84$), DVN$s_1$ ($R=-0.82$) and the energy levels D$E_{2^+}$ ($R=-0.79$), D$E_{4^+}$ ($R=-0.88$), D$E_{6^+}$ ($R=-0.84$). The energy levels usually correlate strongly with each other, and this is inherited by the B(E2)$\uparrow$-s that depend on the $2+$ states. The same is true about the occupancies and vacancies that correlate, passing on their correlation properties to other observables that depend on any of them. 

\section{Statistical inference based on the Bayesian Model Averaging}\label{statistical-inference}

\begin{figure}
\includegraphics[clip, trim=1.5cm 0.5cm 1cm 0.5cm, width=9cm]{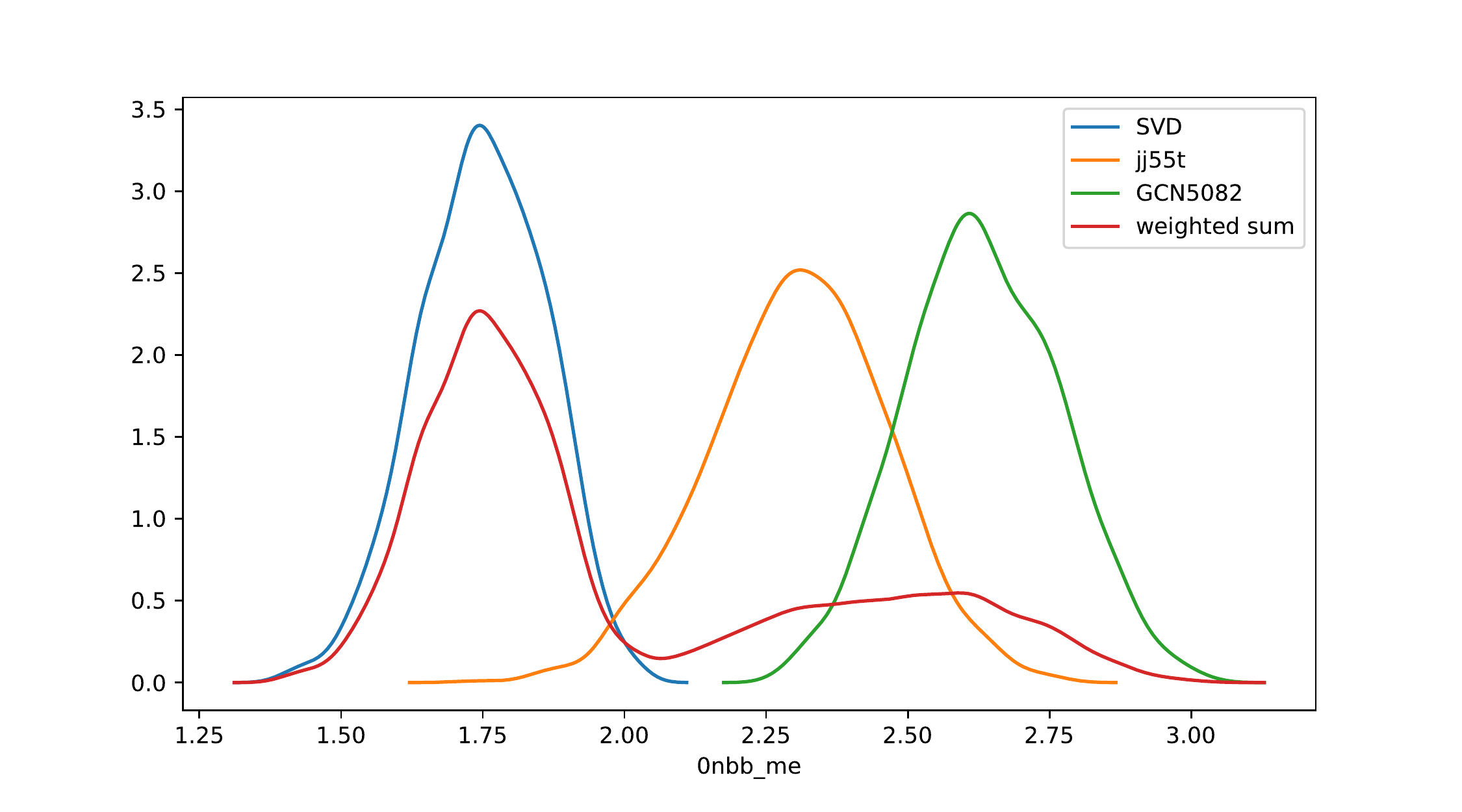}
\caption{PDFs of the $0\nu\beta\beta$ NME distributions for the SVD, jj55t and gcn5082 Hamiltonians, and their weighted sum (see text for details).}
\label{nme-range}
\end{figure}

An in-depth statistical analysis of the $0\nu\beta\beta$ NME may yield valuable insights into its potential range and mean value. It appears that the values of all observables listed in Table \ref{allstats} remain relatively consistent in response to slight variations in the effective Hamiltonian. There is no indication of any significant deviation from the main values, which may be attributed to the preservation of nuclear many-body symmetries in the nuclear shell model. Further investigations, such as utilizing the distributions of available effective Hamiltonians, may aid in determining optimal values and a potential range of error for the $0\nu\beta\beta$ NME. One possible approach investigated in Ref. \cite{Horoi-prc22} for $^{48}$Ca is to combine the distributions for each starting effective Hamiltonian depicted in Fig. \ref{svd-diag-dist} using weighting factors $W_H$,

\begin{equation}
\label{superposition}
\begin{split}
P(x=M_{0\nu})   = & W_{svd} P_{svd}(x) + W_{gcn} P_{gcn}(x) \\
  + & W_{j5t} P_{j5t}(x) \ ,
\end{split}
\end{equation}

where $x$ is the random value of the $0\nu\beta\beta$ NME. The normalized weights $W_k$ with $k=svd,gcn,jj5$ can be inferred 
using the statistical distributions of the evidence observables in Table \ref{allstats} and their correlations with the calculated $M_{0\nu}$ NME.
However, in Ref. \cite{Horoi-prc22} not all data that strongly correlated to the $0\nu\beta\beta$ NME was available, and therefore we only used "democratic" weights for three effective Hamiltonians. In the case of $^{136}$Xe we have all needed data listed in Table \ref{allstats}.
Here, we use the Bayesian Model Averaging method~\cite{Kejzlar_2020,Cirigliano_2022} by re-mapping the quantities in Eq. (\ref{superposition}) in the framework of the Bayes approach to updating probabilities,

\begin{widetext}
\begin{equation} \label{bam}
p(x=M_{0\nu}|y_e,\sigma_e)=\sum_{k=svd}^{j5t} p(x=M_{0\nu}|y_e,\sigma_e,{\cal M}_k) p({\cal M}_k | y_e,\sigma_e) ,
\end{equation}
where $p(x=M_{0\nu}|y_e,\sigma_e,{\cal M}_k)$ correspond to the probability densities $P_k$ in Eq. (\ref{superposition}) and $p({\cal M}_k | (y_e,\sigma_e)$ correspond to the weights $W_k$. Here the ${\cal M}_k$ models are represented by the TBME of different starting Hamiltonians in the jj55 model space, such as SVD, jj55t or gcn5082. The $y_e$ and $\sigma_e$ represent a set of relevant experimental data and their uncertainties for the nuclei involved in the decay. The TBME values for the starting Hamiltonians of each model were determined from a wider set of data (more specific only a set of excitation energies) describing a larger class of nuclei in a given s.p. particle space.  One would like to calibrate the weighting factors $W_k$ to  the evidence data $y_e$ and their errors $\sigma_e$ that are relevant for the $0\nu\beta\beta$ decay, evidence data listed in the first three columns of Table \ref{allstats}. In Eq. (\ref{bw}), $\theta_j$ represents a set of parameters describing the model ${\cal M}_k$ Hamiltonians, i.e. their two-body matrix elements. To obtain the weights $W_k$ one needs the so called evidence integrals

\begin{equation} \label{bw}
p(y_e,\sigma_e|{\cal M}_k) = \int \prod_i^{N_{obs}} dy_i p_{y_e,\sigma_e}(y_i)\left[\int \prod_j^{N_{tbme}} d\theta_j p(y_i|\theta_j, {\cal M}_k) \pi (\theta_j|{\cal M}_k) \right] ,
\end{equation}

\end{widetext}
which can be used in Bayes theorem to obtain the posterior probabilities

\begin{equation}
p({\cal M}_k| y_e,\sigma_e) = \frac{p(y_e,\sigma_e|{\cal M}_k) \pi({\cal M}_k)}{\sum_{k=svd}^{j5t} p(y_e,\sigma_e|{\cal M}_k) \pi({\cal M}_k)} .
\end{equation}
Here the $\pi({\cal M}_k)$ are the prior probabilities for each model, which are considered uniform. In Eq. (\ref{bw}) $\pi (\theta_j|{\cal M}_k)$ represents the distribution of the parameters $\theta_j$ in a given model, which we generate uniformly, although with a compact support. In addition, the evidence likelihood function is taken as a typical product for independent observables,
\begin{equation}
 p_{y_e,\sigma_e}(y_i) \propto \prod_i^{N_{obs}}  exp[-(y_i - y_{e_i})^2/(2\sigma_{e_i}^2)] ,
\end{equation}
where the overall proportionality factor is irrelevant if the same set of observales are used with all models. In Eqs. (\ref{bam}-\ref{bw}) the  integration variable $y_i$ run on a subset of observables that correlate strongly with $M_{0\nu}$. Here we take the 10 observables that have $R>0.5$, listed below $0\nu\beta\beta$ in the heat map of Fig. \ref{svd-hm} and included in the correlation matrix of Fig. \ref{svd-corr}. The integrals in eq. (\ref{bw}) can be done using Monte Carlo techniques, provided that the integration hypervolume is the same for all models ${\cal M}_k$.
Having the  posterior probabilities $p({\cal M}_k| y_e,\sigma_e)$, one often calculates the Bayesian factors
\begin{equation}
B^k_m = p({\cal M}_k| y_e,\sigma_e)/p({\cal M}_m| y_e,\sigma_e)
\end{equation}
to either infer that one model is dominant or to use them in Eq. (\ref{superposition}) (equivalent of Eq. (\ref{bam})) to obtain an average probability density. In our case, using a standard quenching factor of 0.7 for all GT matrix elements one gets a clearly dominant contribution of the SVD model. In principle, one could infer that all $W_k$ are 0, but the $W_{svd}$. However, given the inherent bias embedded in the Bayesian approach, and in the spirit of the predictor-corrector approach to step-by-step evolution schemes, we consider for the weights $W_k$ an average between the prior probabilities $\pi({\cal M}_k)$ and the posterior probabilities $p({\cal M}_k| y_e,\sigma_e)$.

Fig. \ref{nme-range} shows the probability distribution functions (PDF) for the three starting effective Hamiltonians and their weighted sum. To calculate each PDF we use kernel-density estimates \cite{Rosenblatt_1956,Silverman-kde} for the histograms describing the $M_{0\nu}$, such as that of the upper-left panel of Fig. \ref{svd-diag-dist}. Based on the results of our statistical analysis summarized in Fig. \ref{nme-range} (see the "weighted sum" curve) one can infer that with 90\% confidence the $0\nu\beta\beta$ NME lies in the range between 1.55 and 2.65, with a mean value of about 1.99 and a standard deviation of 0.37. 

The results presented above may vary if a different quenching factor, $q$, is used. In this study, $q=0.7$ was employed, which aligns with values used in commonly studied model spaces such as $sd$, $fp$, and $jj44$ \cite{BrownRichter2006,Horoi-prc22,JUN45}. For instance, if a very low quenching factor, less than 0.45, is used, the impact of the GCN5082 effective Hamiltonian in Eq. (\ref{superposition}) will increase, resulting in a shift of the weighted distribution in Fig. \ref{nme-range} towards higher values. This correlation between the $0\nu\beta\beta$ NME and the $2\nu\beta\beta$ NME, and the need for a small quenching factor, $q \approx 0.35$, to describe the $2\nu\beta\beta$ NME could be an effect of excessive adjustments to the TBME of the effective Hamiltonian due to fine-tuning the energies~\cite{BrownRichter2006}. Similarly, when $0.45 < q < 0.56$, the distribution for the $jj55t$ dominates the weighted distribution in Fig. \ref{nme-range}.

\section{Conclusion and Outlook} \label{conclusions}

In this paper we presented a statistical model for analyzing the distribution and the theoretical uncertainty of the $0\nu\beta\beta$ NME of experimentally relevant isotope $^{136}$Xe (see e.g. Ref. \cite{PhysRevLett.130.051801} for the latest $0\nu\beta\beta$ lower half-life limit), using the interacting shell model in the $jj55$-shell model space. For this analysis we used three known starting effective Hamiltonians that were widely tested for tin isotopes and other nuclei near $^{132}$Sn, namely SVD, GCN5082 and jj55t.  Considering potential uncertainties of these starting effective Hamiltonians, we added to their TBME random contributions of $\pm$10\%. Using sample sizes of 1,000 points we analyzed
for each starting effective Hamiltonian: (i) the correlations between $0\nu\beta\beta$ NME and other 23 observables that are accessible experimentally; (ii) the theoretical ranges for each observables; (iii)  the shape of different distributions for each observables and each starting Hamiltonian; (iv) the weighted contributions from different starting Hamiltonians to the "optimal" distribution of the $0\nu\beta\beta$ NME; (v) an "optimal" value of the $0\nu\beta\beta$ NME and its predicted probable range (theoretical error).

As in the case of $^{48}$Ca \cite{Horoi-prc22}, we found that the $0\nu\beta\beta$ NME correlates strongly with the $2\nu\beta\beta$ NME, but much less with the Gamow-Teller strengths to the first $1^+$ state in $^{136}$Cs. We also found that the $0\nu\beta\beta$ NME exhibits reasonably strong correlations with the energies of the $2^+$, $4^+$ and $6^+$ states in $^{136}$Ba, and with the neutron occupation probabilities in $^{136}$Xe.
We also noticed that there are additional correlations between observables, such as the energies of the $2^+$, $4^+$ and $6^+$ states in $^{136}$Ba and the neutron occupation probabilities, as well as between B(E2)$\uparrow$ values in $^{136}$Ba and proton and neutron occupation probabilities, which can indirectly influence the $0\nu\beta\beta$ NME. 

The significant difference in the present analysis relative to that for $^{48}$Ca \cite{Horoi-prc22} is that reliable experimental values for the occupation probabilities in $^{136}$Ba and $^{136}$Xe are available. This made possible a statistical analysis of the $0\nu\beta\beta$ NME within the Bayesian Averaging Model that can all the available experimental data for the nuclei involved in the transition to update the weights corresponding to each starting Hamiltonian to the overall NME distribution. Based on this statistical analysis with three independent starting effective Hamiltonians we propose a common probability distribution function for the $0\nu\beta\beta$ NME, which has a range (theoretical error)  of (1.55 - 2.65) at 90\% confidence level, with a mean value of 1.99 and a standard deviation of 0.37. 

Unfortunately, our results still depend on the choice of the quenching factor for the Gamow-Teller operator. Ab-initio studies, however, can overcome this shortcoming by consistently producing effective operators that can describe Gamow-Teller transition without the need of a quenching factor. We thus believe that the present analysis will help ab-initio studies, such as those reported in Refs.\cite{Yao2020,Novario2021,Belley2021}, to better identify correlations and further reduce the uncertainties of the $0\nu\beta\beta$ NME, given that such ab-initio analyses, e.g. that recently reported on $^{76}$Ge~\cite{Belley-2022}, seem to be confined to fewer observables and much smaller statistics.

\vspace{0.5cm}
    {\it Acknowledgements}. MH acknowledges support from the US Department of Energy grant DE-SC0022538 "Nuclear Astrophysics and Fundamental Symmetries". %{\color{teal} 
    AN and SS acknowledge support by grants of Romanian Ministry of Research, Innovation and Digitalization through the project CNCS – UEFISCDI number 99/2021 within PN-III-P4-ID-PCE-2020-2374 and the project CNCS – UEFISCDI number TE12/2021 within PN-III-P1-1.1-TE-2021-0343. We are grateful for the resources at INCDFM-CIFRA HPC Cluster.

\bibliographystyle{apsrev4-2} %de reactivat
\bibliography{bb-n}

\end{document}